\documentclass[12pt,english,aps,manuscript]{revtex4-1}
\usepackage[T1]{fontenc}
\usepackage[latin9]{inputenc}
\usepackage[letterpaper]{geometry}
\usepackage{array}
\usepackage{longtable}
\usepackage{units}
\usepackage{multirow}
\usepackage{amsmath}
\usepackage{amssymb}
\usepackage{graphicx}
\usepackage{comment}
\usepackage{esint}

\makeatletter

\providecommand{\tabularnewline}{\\}

\@ifundefined{textcolor}{}
{%
 \definecolor{BLACK}{gray}{0}
 \definecolor{WHITE}{gray}{1}
 \definecolor{RED}{rgb}{1,0,0}
 \definecolor{GREEN}{rgb}{0,1,0}
 \definecolor{BLUE}{rgb}{0,0,1}
 \definecolor{CYAN}{cmyk}{1,0,0,0}
 \definecolor{MAGENTA}{cmyk}{0,1,0,0}
 \definecolor{YELLOW}{cmyk}{0,0,1,0}
 }

\usepackage{mathrsfs}
\usepackage{bbm}
\usepackage{babel}
\usepackage{textcomp}

\makeatother

\usepackage{babel}

\begin{document}

\title{Implementation of quantum logic gates using  polar molecules in pendular states}

\author{Jing Zhu and Sabre Kais}

\thanks{Corresponding author, kais@purdue.edu}

\affiliation{Department of Chemistry, Physics  and Birck Nanotechnology Center, Purdue University,
West Lafayette, IN 47907, USA}

\author{Qi Wei and Dudley Herschbach}

\affiliation{Department of Physics, Texas A$\&$M University, College Station,
Texas 77843, USA}

\author{Bretislav Friedrich}

\affiliation{Fritz-Haber-Institut der Max-Planck-Gesellschaft, Faradayweg 4-6,
D-14195 Berlin, Germany}

\begin{abstract}

We present a systematic approach to implementation of basic quantum logic gates operating on polar molecules in pendular states as qubits for a quantum computer. A static electric field prevents quenching of the dipole moments by rotation, thereby creating the pendular states; also, the field gradient enables distinguishing among qubit sites. Multi-Target Optimal Control Theory (MTOCT) is used as a means of optimizing the initial-to-target transition probability via a laser field. We give detailed calculations for the SrO molecule, a favorite candidate for proposed quantum computers. Our simulation results indicate that NOT, Hadamard and CNOT gates can be realized with high fidelity, as high as 0.985,  for such pendular qubit states.

%We consider also a recent proposal to use of polar symmetric top molecules in quantum computing. That offers advantages resulting from a first-order Stark Effect which renders the effective dipole moments nearly independent of the field strength. For a particular choice of qubits, the electric dipole interactions then become isomorphous with NMR systems.  Thereby, the logic gate operations can be enhanced in analogy to NMR quantum computers.

\end{abstract}
\maketitle

\section{Introduction}

Quantum computers take direct advantage of superposition and entanglement to perform computations.
Because quantum algorithms compute in ways which classical computers cannot,
for certain problems they provide exponential
speedups over their classical counterparts
\cite{Micheli2006,Ni2008a,Kotochigova2006a,Deiglmayr2008,DeMille2002,Carr2009,Anmer,zhuminima}.
That prospect has fostered a variety of proposals suggesting means to implement such a device
\cite{DeMille2002, Sorensen2004, Wallraff2004}.
DeMille has detailed a prototype design for quantum computation using ultracold polar molecules, trapped in a one-dimensional optical lattice, partially oriented in an external electric field, and coupled by the dipole-dipole interaction \cite{DeMille2002}.
This offers a promising platform for quantum computing because scale-up appears feasible to obtain large networks of coupled qubits \cite{Carr2009, Friedrich2009, Lee2005, Kotochigova2006a, Yelin2006, Ni2008a}.

In previous work, we focused on entanglement and on consequences of using a strong external electric field with appreciable gradient, required to prevent quenching of the dipole moments by rotation and to enable addressing individual qubit sites \cite{QiPendular}. The molecules were represented as identical, rigid dipoles a fixed distance apart and undergoing pendular oscillations imposed by the external electric field. We determined the dependence of the entanglement of the pendular qubit states, as measured by the concurrence function, on three unitless variables, all scaled by the rotational constant.
The first specifies the Stark energy and intrinsic angular shape of the qubits; the second specifies the magnitude of the dipole-dipole coupling; the third variable specifies the thermal energy.
Under conditions deemed amenable for proposed quantum computers, we found that both the concurrence and a key frequency shift, $\Delta \omega$, that has a major role in logic gates, become very small for the ground eigenstate. In order that such weak entanglement can suffice for operation of logic gates, the resolution must be high enough to detect the $\Delta \omega$ shift unambiguously.

For diatomic molecules, the Stark effect is second-order, therefore a sizable external electric field is required to produce the requisite dipole moments in the laboratory frame. In a subsequent study, we examined symmetric top molecules as candidate qubits \cite{symmetricTop}. Symmetric top molecules offer advantages resulting from a first-order Stark effect, which renders the effective dipole moments nearly independent of the field strength. That permits the use of a much lower external field strength in addressing sites. Moreover, for a particular choice of qubits, the electric dipole interactions become isomorphous with NMR systems.

Here we study further aspects of how to implement a set of basic quantum gates for pendular qubit states of polar diatomic or linear molecules.
We apply the Multi-Target Optical Control Theory (MTOCT) \cite{Vivie1999,Vivie2004,Vivie2005,Vivie2011,Rabitz2000,Rabitz2010,Yamashita2009a,Yamashita2009b,Yamashita2010,Yamashita2011}
 to design laser pulses that enable resolving and inducing transitions between specified states of the qubit system.
 This approach has been previously employed to study optimal control for elements of quantum computation in molecular systems, using as qubits vibrational or rotational states \cite{Yamashita2009b,Bomble2010,Vivie2011,Yamashita2010,Sugny2009}.
 Our use of pendular qubit states incorporates more fully effects of the external electric field and thereby simplifies the gate operations.

Section II specifies the Hamiltonian defining the pendular qubits, as well as fundamental aspects of MTOCT.
In Sec. III we present simulation results using MTOCT to obtain optimized laser pulses for realizing NOT, Hadamard and CNOT logic gates;
 those gates with the  addition of the phase gate $\pi/8$ provide the basis for universal quantum computation \cite{Book_Quantum}.
Section IV discusses strategies to contend with cases in which the $\Delta \omega$ shift is zero or becomes too small to resolve.

\section{Theory}

\subsection{Eigenstates for Polar Molecules in Pendular States}

The Hamiltonian for an individual trapped polar diatomic or linear molecule in an external electric field
$\boldsymbol{\epsilon}$ can, for our purposes, be reduced just to the rotational kinetic energy and Stark interaction terms \cite{QiPendular}:

\begin{equation}
\mathcal{H}_{S}=B\cdot\boldsymbol{J}^{2}-\mu\epsilon\cos\theta
\label{eq:reduced}
\end{equation}

This represents a spherical pendulum: $B\boldsymbol{J}^{2}$ is the rotational energy, with $B$ the rotational constant, and $\mu$ the permanent dipole moment; $\theta$ is the polar angle between the molecular axis and the external field direction. At the ultracold temperatures that we consider, the translational kinetic energy of the trapped molecules is very small and nearly harmonic within the trapping well, so the trapping energy is nearly constant and hence is omitted.    Interactions involving open shell electronic structure or nuclear spins or quadrupole moments are also omitted (but could be incorporated in familiar ways\cite{PRL109}). The eigenstates of $\mathcal{H}_{S}$, resulting from mixing of the field-free rotational states by the Stark interaction, are designated as pendular states.   As proposed by DeMille \cite{DeMille2002}, the qubits $\left|0\right\rangle $ and $\left|1\right\rangle $  are chosen as the two lowest $M=0$ pendular states, with  $\tilde{J}=0$ and $1$, respectively.  Here the $\tilde{J}$ notation (tilde-hat) indicates it is no longer a good quantum number, due to the Stark mixing.  However, $M$ remains good as long as azimuthal symmetry about $\boldsymbol{\epsilon}$ is maintained. The qubits thus are superpositions of spherical harmonics,

\begin{equation}
\left|0\right\rangle =\sum_{j}a_{j}\cdot Y_{j,\,0}\left(\theta,\,\varphi\right);\;\;\;
\left|1\right\rangle =\sum_{j}b_{j}\cdot Y_{j,\,0}\left(\theta,\,\varphi\right)
\label{eq:qubits definition}
\end{equation}

\begin{figure}
\begin{centering}
\includegraphics[width=0.7\textwidth]{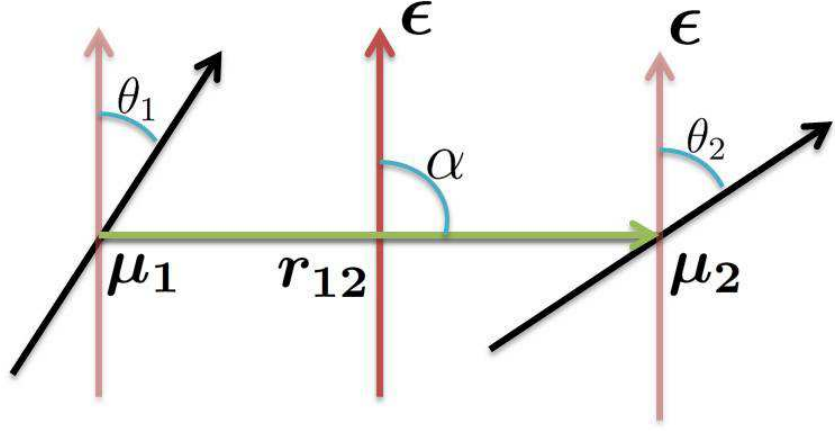}
\par\end{centering}

\caption{The configuration of two polar molecules. $\boldsymbol{\mu_{1}}$
and $\boldsymbol{\mu_{2}}$ are permanent dipole moments for molecule $1$
and $2$; $\boldsymbol{r_{12}}$ is the distance vector from molecule $1$
to $2$. $\boldsymbol{\epsilon}$ is the external electric field and $\alpha$
is the angle between $\boldsymbol{r_{12}}$ and the external field. $\theta_{1}$
($\theta_{2}$) is the angle between $\boldsymbol{\mu_{1}}$ ($\boldsymbol{\mu_{2}}$)
and $\boldsymbol{\epsilon}$. \label{skewmap}}
\end{figure}

Adding a second molecule into the trap, identical to the first but at distance $r_{12}$ from it, introduces in addition to its pendular term the dipole-dipole interaction, $V_{dd}$.  Averaging over the azimuthal angles  \cite{QiPendular}, which for $M = 0$ states are uniformly distributed, reduces the dipole-dipole interaction to

\begin{equation}
V_{dd}=\Omega_{\alpha}\cdot\cos\theta_{1}\cdot\cos\theta_{2}
\label{eq:vdd}
\end{equation}

where $\Omega_{\alpha}=\Omega\left(1-3\cos^{2}\alpha\right)$, with $\Omega=\nicefrac{\mu^{2}}{r_{12}^{3}}$.
As depicted in Fig. \ref{skewmap}, $\alpha$ is the angle between the array axis and the electric field direction $\boldsymbol{\epsilon}$;
 $\theta_{1}$ and $\theta_{2}$ are the polar angles between the dipoles and the field direction.

To exemplify logic gate operations, it is sufficient to consider just two molecules.
In the basis set of qubit pendular states
$\left\{ \left|00\right\rangle ,\,\left|01\right\rangle ,\,\left|10\right\rangle ,\,\left|11\right\rangle \right\} $,
the two-molecule Hamiltonian can be expressed as
$\mathcal{H}_{tot}=\mathcal{H}_{S1}+\mathcal{H}_{S2}+V_{dd}$, with

\begin{eqnarray}
\mathcal{H}_{S1} & = & \left(\begin{array}{cc}
W_{0}\\
 & W_{1}
\end{array}\right)\otimes\mathbf{I}_{2};\;\mathcal{H}_{S2}=\mathbf{I}_{2}\otimes\left(\begin{array}{cc}
W_{0}^{\prime}\\
 & W_{1}^{\prime}
\end{array}\right)\nonumber \\
V_{dd} & = & \Omega_{\alpha}\left[\left(\begin{array}{cc}
C_{0} & C_{x}\\
C_{x} & C_{1}
\end{array}\right)\otimes\left(\begin{array}{cc}
C_{0}^{\prime} & C_{x}^{\prime}\\
C_{x}^{\prime} & C_{1}^{\prime}
\end{array}\right)\right]
\label{eq:vdd1}
\end{eqnarray}

Here $W_{0}$ and $W_{1}$ are eigenenergies of the pendular qubits
for states $\left|0\right\rangle $ and $\left|1\right\rangle $.
$\mathbf{I}_{2}$ is the $2\times2$ identity matrix.
$C_{0}$, $C_{1}$ are the expectation values of $\cos\theta$ in the basis of
$\left|0\right\rangle$ and $\left|1\right\rangle$;
while $C_{x}$ indicates the transition dipole moment between $\left|0\right\rangle$ and $\left|1\right\rangle$.
The matrix elements thus are defined as:

\begin{equation}
C_{0}=\left\langle 0\left|\cos\theta\right|0\right\rangle ;\;\;
C_{1}=\left\langle 1\left|\cos\theta\right|1\right\rangle ;\;\;
C_{x}=\left\langle 0\left|\cos\theta\right|1\right\rangle
\label{eq:c01x}
\end{equation}

In Eq. \ref{eq:vdd}, primes are used to indicate that the external field strength differs at the location of the two molecules, as required to distinguish between the qubit sites.  The entanglement of the pendular qubit states of $H_{tot}$ was evaluated in ref \cite{QiPendular}. From numerical results, simple approximate formulas were obtained that provide the concurrence and the key frequency shift, $\Delta \omega$, in terms of two unitless reduced variables:
$x=\nicefrac{\mu \epsilon}{B}$ and $\nicefrac{\Omega_{\alpha}}{B}$. When $\nicefrac{\Omega_{\alpha}}{B} \ll 1$, the usual case, the concurrence is proportional to $\Delta \omega$, which is given by
\begin{equation}
\Delta \omega = \left|\Omega_{\alpha} \right| \left(C_{1}-C_{0}\right)\left(C^{\prime}_{1}-C^{\prime}_{0}\right)
\label{eq:deltaomega}
\end{equation}

\begin{figure}
\begin{centering}
\includegraphics[width=0.9\textwidth]{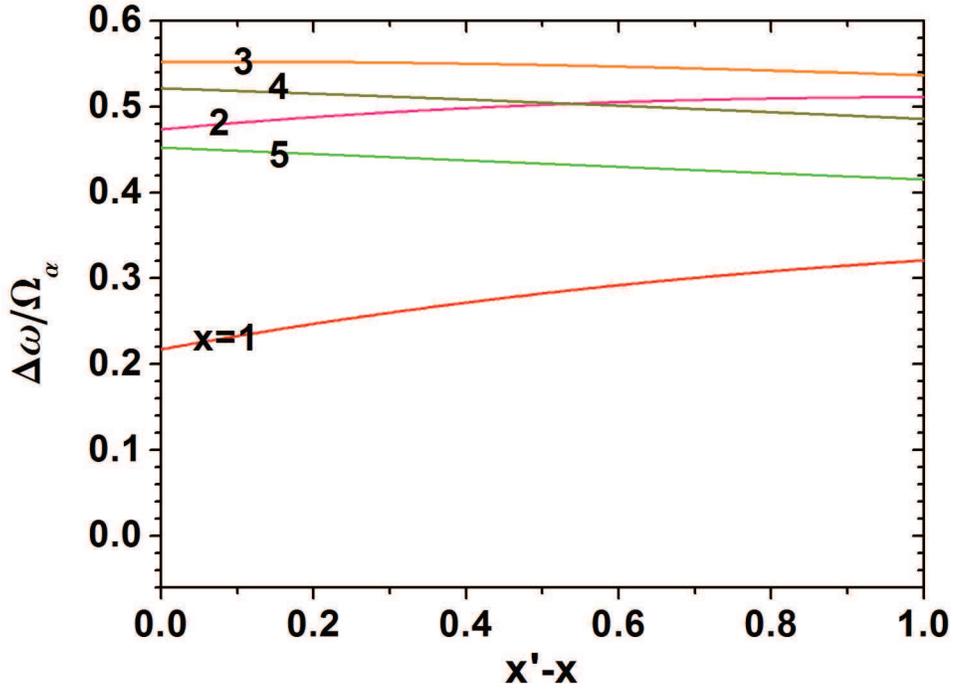}
\par\end{centering}

\caption{Ratio of frequency shift, $\Delta \omega$, to the dipole-dipole interaction parameter,
$\Omega_{\alpha}= \left(1 - 3\cos^{2}\alpha\right) \mu ^{2}/r_{12}^{3}$, as a function of reduced field strength,
$x = \nicefrac{\mu \boldsymbol{\epsilon}}{B}$ and $x^{\prime} = \nicefrac{\mu \boldsymbol{\epsilon}^{\prime}}{B}$ at sites of the two dipoles. \label{ratio}}
\end{figure}

Figure \ref{ratio} plots this relation, wherein $\nicefrac{\Delta \omega}{\Omega_{\alpha}}$ depends only on $x$ and $x^{\prime}-x$.

\subsection{Multi-Target Optical Control Theory (MTOCT)}

Much attention has recently been devoted to applying Optimal Control Theory
 for elements of quantum  computation in molecular systems\cite{Vivie1999, Vivie2004, Vivie2005, Vivie2011}.
 The basic idea is to design laser pulses which allow manipulation of
transitions within each qubit separately. For implementing basic quantum gates,
the aim is to achieve large transition probabilities with the correct phase from
a specific initial state into a final target state by application of an external laser field
while minimizing the laser energy. For our case, we can construct the following
 MTOCT objective function, $\Im$, which needs to be maximized~\cite{Yamashita2009b}:

\begin{multline}
\Im\left[\psi_{ik}\left(t\right),\,\psi_{fk}\left(t\right),\,\epsilon\left(t\right)\right]
=\sum_{k=1}^{z}\left\{ \left|\left\langle \psi_{ik}\left(T\right)|\phi_{fk}\right\rangle \right|^{2}
\vphantom{\intop_{0}^{T}\left\langle \psi_{fk}\left(t\right)\left|i\left[H-\boldsymbol{\mu}\cdot\boldsymbol{E}\left(t\right)\right]+\frac{\partial}{\partial t}\right|\psi_{ik}\left(t\right)\right\rangle dt}\right.
-\alpha_{0}\intop_{0}^{T}\frac{\left|\boldsymbol{E}\left(t\right)\right|^{2}}{S\left(t\right)}dt-2Re\left\{ \left\langle \psi_{ik}\left(T\right)|\phi_{fk}\right\rangle\right.\\
\left.\left.\times\intop_{0}^{T}\left\langle \psi_{fk}\left(t\right)\left|\frac{i}{\hbar}\left[H-\boldsymbol{\mu}\cdot\boldsymbol{E}\left(t\right)\right]+\frac{\partial}{\partial t}\right|\psi_{ik}\left(t\right)\right\rangle dt\right\} \right\}
\label{eq:target function}
\end{multline}

where  $z$ is the total
number of targets and for $N$ qubits  is given by $z=2^{N}+1$,
where $2^N$ is the number of input-output transitions in the gate transformation and the supplementary equation is the phase constraint.
Thus, for the two dipole system, $z=5$.
Here, $\psi_{ik}$ is the wave function of the $k$-th target, driven by the laser field
$E\left(t\right)$, with initial condition $\psi_{ik}\left(0\right)=\phi_{ik}$.
Whereas $\psi_{fk}$ is the wave function of the $k$-th target driven
by the same laser with final condition $\psi_{fk}\left(T\right)=\phi_{fk}$. Thus, the first term on the right-hand side represents the overlap between
the laser driven wavefunctions and the desired target states.  In the second term,
$E\left(t\right)$ is the laser intensity,
$S\left(t\right)=sin^{2}\left(\nicefrac{\pi t}{T}\right)$ is the
laser envelop function, which guarantees the experimentally appropriate slow turn-on and turn-off laser pulse envelope \cite{Vivie1999,PRL108}.
$T$ is the total duration time of the laser.
$\alpha_{0}$ is a positive penalty factor chosen to weight the importance of the laser fluence.  The last term  denotes the time-dependent Schr\"odinger equations for wave functions
$\psi_{ik}(t)$ and $\psi_{fk}(t)$ with $\mathcal{H}_{tot}$ the Hamiltonian of the system.

Requiring $\delta \Im=0$, specifies the equations satisfied by the wave function, the Lagrange multiplier, and the optimized laser field \cite{Yamashita2009b}:
\begin{eqnarray}
i\hbar\frac{\partial}{\partial t}\psi_{ik}\left(t\right) & = & \left\{ \mathcal{H}_{tot}-\boldsymbol{\mu}\cdot\boldsymbol{E}\left(t\right)\right\} \psi_{ik}\left(t\right)\nonumber \\
i\hbar\frac{\partial}{\partial t}\psi_{fk}\left(t\right) & = & \left\{ \mathcal{H}_{tot}-\boldsymbol{\mu}\cdot\boldsymbol{E}\left(t\right)\right\} \psi_{fk}\left(t\right)\nonumber \\
 &  & \psi_{ik}\left(0\right)=\phi_{ik};\;\psi_{fk}\left(T\right)=\phi_{fk}
 \label{eq:schrodinger}\\
 &  & and\; k=1\cdots z\nonumber
\end{eqnarray}

\begin{eqnarray}
E\left(t\right) & = & -\frac{z\cdot \mu \cdot S\left(t\right)}{\hbar\cdot\alpha_{0}}\cdot\sum_{k=1}^{z}\mbox{Im}\left\{ \left\langle \psi_{ik}\left(t\right)|\psi_{fk}\left(t\right)\right\rangle \right.\nonumber \\
 &  & \left.\left\langle \psi_{fk}\left(t\right)\left|cos\theta_{1}+cos\theta_{2}\right|\psi_{ik}\left(t\right)\right\rangle \right\}
 \label{eq:Et}
\end{eqnarray}

In order to examine the performance of the optimized laser pulse, we evaluate two factors \cite{Yamashita2009b,Palao2003,Shioya2007}:
the average transition probability, given by,

\begin{equation}
\bar{P}=\frac{1}{z}\cdot\sum_{k=1}^{z}\left|\left\langle \psi_{ik}\left(T\right)\right|\left.\Phi_{fk}\right\rangle \right|^{2}
\label{eq:averagetransion}
\end{equation}

and the fidelity, given by
\begin{equation}
F=\frac{1}{z^{2}}\cdot\left|\sum_{k=1}^{z}\left\langle \psi_{ik}\left(T\right)\right|\left.\Phi_{fk}\right\rangle \right|^{2}
\label{eq:fidelity}
\end{equation}

The average transition probability involves only the overlap
of $\psi_{ik}\left(T\right)$ and $\Phi_{fk}$, but does not reflect
the difference of phase information between the laser driven final
state $\psi_{ik}\left(T\right)$ and the designed target state $\Phi_{fk}$.
Since the phase information is very important for quantum logical gates,
we use only the fidelity parameter to assess our simulation results.

\section{Simulation Results for Polar Diatomic Molecules}

A number of diatomic polar molecules offer properties suitable for a quantum computer \cite{DeMille2002,Lee2005,Yelin2006,QiPendular,Yamashita2009b,Yamashita2011}.  For our numerical study, we chose SrO, for which the dipole moment $\mu=8.9$ Debye and rotational constant $B = 0.33\,\unit{cm^{-1}}$ \cite{SrO}.  Since we consider trap temperatures in the microkelvin range, with $\nicefrac{k_{B}T}{B}\,\sim\,10^{-6}$, thermal excitations are negligible. To specify the pendular states and other properties requires assigning the external field strengths at the sites of the two molecules and the distance between them.  We used field strengths (in scaled units) of $\mu \boldsymbol{\epsilon} = 2$ and 3, corresponding to $\boldsymbol{\epsilon}=4.4$ and 6.6 kV/cm, respectively, at the two sites. Initially, we took $r_{12} = 500\,\unit{nm}$, a typical spacing for molecules trapped in an optical lattice \cite{DeMille2002,QiPendular}.  We set the angle $\alpha\,=\,90^{o}$ (cf. Fig. \ref{skewmap}), the usual experimental choice.    Then $\nicefrac{\Omega_{\alpha}}{B}=9.7\times10^{-5}$. The corresponding pendular eigenstate reduced energies are $\nicefrac{E_{i}}{B} = -1.65,\,1.19,\, 1.92,\, 4.77$ and the cosine matrix elements are $C_{0} = 0.480$, $C_{1} = -0.208$; $C_{0}^{\prime} = 0.579$ and $C_{1}^{\prime} = -0.164$. Thus, from Eq. \ref{eq:deltaomega} we obtain $\Delta \omega = 51 \; \unit{kHz}$.  As seen in Fig. \ref{ratio},  $\Delta \omega$ varies only modestly with the field strength in the range
$\nicefrac{\mu \boldsymbol{\epsilon}}{B}= 2 - 5$, considered optimum \cite{DeMille2002,QiPendular}, but  $\Delta \omega$ is directly proportional to the dipole-dipole coupling strength, $\Omega_{\alpha}$.

At present, it remains an open question whether, in the presence of line broadening induced by the static external electric field, adequate resolution can be obtained to resolve unambiguously a frequency shift of only ~50 kHz \cite{QiPendular}. Such a small $\Delta \omega$ is also a severe handicap for our theoretical simulation of laser-driven logic gates. For instance, the laser pulse duration \cite{Bomble2010} required for realizing the CNOT gate is $\tau\,=\,\nicefrac{10\hbar}{\Delta \omega}$. Hence a frequency shift so small as 50 kHz requires that the laser pulse duration is at least 31$\unit{\mu s}$.  Recently, Zaari and Brown studied the effect of laser pulse shaping parameters about the fidelity in realizing the quantum gates. They proved that the amplitude variation and frequency resolution plays the important role in the fidelity \cite{Brown2012}. We will explore the impacts of those two coefficients in the future work. That is much longer than the self-evolution period of the system, about 29 ps. If we set the computation time step at 0.25 ps, a single simulation run would need $1.25\times10^{8}$ steps. Our MTOCT calculations involve many iterative runs; e. g., for a CNOT gate about 440 iterations.  With current computers, such a calculation would be daunting: it would need about 700 GB of RAM storage to perform and each iteration would take about 2.8 days, in total nearly 3.5 years for standard i7 core.

To make the calculation feasible, we reduced the spacing between the dipoles ten-fold, which increases $\Omega$ by 1000-fold, and thus $\Delta \omega= 50\,\unit{MHz}$ and the laser pulse duration shortens to ~33 ns.  Then simulation runs have $\sim 10^{5}$ steps, the RAM needed shrinks to 700 MB, and the computation time to about 4 minutes per iteration, so $\sim30$ hours for 440 iterations.  The reduced dipole-dipole spacing, which becomes only 50 nm, actually corresponds to the range recently proposed for plasma-enhanced, electric/electrooptical traps, for which the trap frequencies can exceed 100 MHz \cite{kais2010, Chang2009, Murphy2009}, and might be attainable in an optical ferris wheel device \cite{Franke2007}. Reducing the spacing so markedly is not considered practical, however, because it would strongly foster inelastic, spontaneous Raman scattering of lattice photons and hence induce unacceptably large decoherence \cite{DeMille2002,Carr2009,lattice}. The resort to taking $r_{12}$ unrealistically small was done reluctantly but enables us to illustrate the general utility of MTOCT applied to quantum logic gates.

In the simulations, the time evolution of $\Psi_{ik}\left(t\right)$ and $\Psi_{fk}\left(t\right)$ is calculated from Eq. \ref{eq:schrodinger} by the fourth-order Runge-Kutta method, using time steps of 0.25 ps. The penalty factor $\alpha_{0}$ is set as $5\times10^{6}$, the same as in Ref. \cite{Yamashita2009b}. The penalty factor is used to minimize the fluence of the external fields \cite{Yamashita2011}. For the optimized laser pulse E(t) from Eq. \ref{eq:Et}, we adopted a rapidly convergent iteration using a first-order split-operator approach \cite{Rabitz1998}. The maximum iteration number was set at 600. In addition to the usual four-qubit basis set, $\left\{ \left|00\right\rangle ,\,\left|01\right\rangle ,\,\left|10\right\rangle ,\,\left|11\right\rangle \right\}$, we also included the phase correction, introduced into the MTOCT approach by Tesch and de Vivie-Riedle \cite{Vivie2004}. In ref. \cite{Yamashita2009b}, Mishima and Yamashita pointed out that the purpose of the phase constraint is preventing each state evolving to different phases, which can provide the correct quantum logical gates. Recently, Zaari and Brown pointed out that align the phase of all qubits appropriately can lead to effective subsequent quantum gates (laser pulses) \cite{Brown2011}.

\begin{figure}
\begin{centering}
\includegraphics[width=0.9\textwidth]{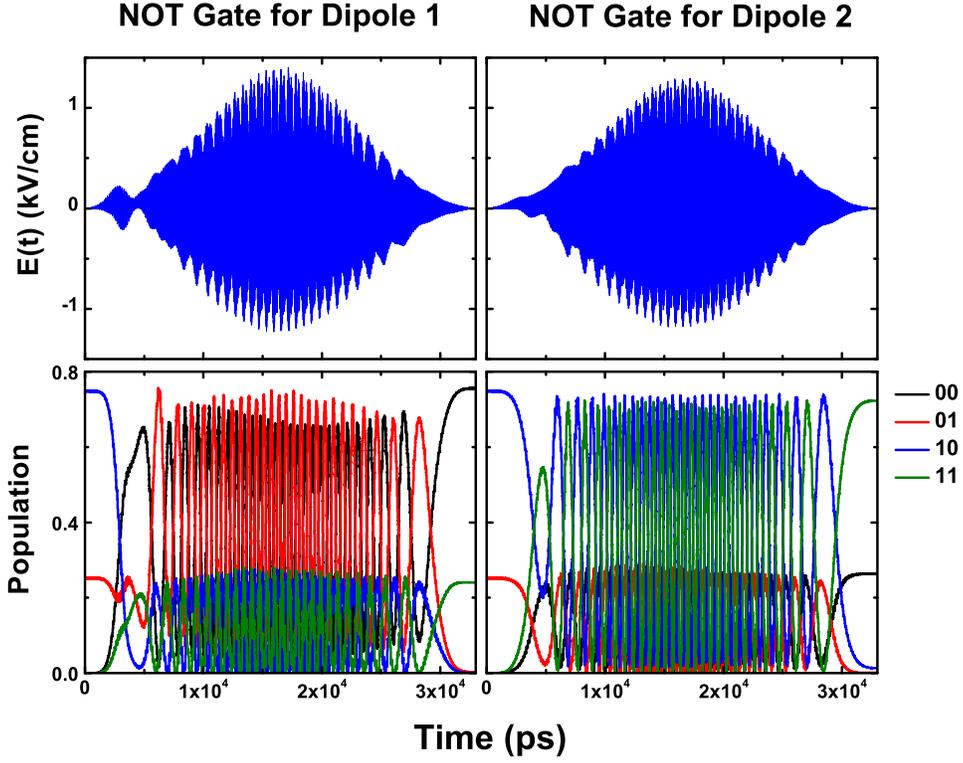}
\par\end{centering}

\caption{Converged laser pulses for NOT gate. The upper panel is the laser pulses
for realizing the NOT gate for dipole $1$, on the left and
 dipole $2$ on the right. The initial and target states
are listed in Table \ref{table_not}.
The lower panel shows the evolution of all populations driven by NOT pulses.
The left panel exhibits the population evolution under the NOT pulse for dipole $1$.
The initial state is
$\cos\left(\frac{\pi}{3}\right)\left|01\right\rangle +\sin\left(\frac{\pi}{3}\right)\left|10\right\rangle $
and the final populations
are $0.733$ for $\left|00\right\rangle $ and $0.265$ for $\left|11\right\rangle $.
The right panel shows the population evolution via the NOT pulse for dipole
$2$. It has the same initial condition as the  left  panel and the
converged population is $0.216$ for $\left|00\right\rangle $ and
$0.781$ for $\left|11\right\rangle $. Both pulses perform the corresponding
NOT gates nicely.
 \label{NOT}}
\end{figure}

In Table \ref{table_not}, we show both the initial and target states for the NOT gate for the two dipoles. The optimized laser pulse for NOT gates applied to dipole 1 and dipole 2 separately is shown in the upper panel of Fig. \ref{NOT}. The pulse for the NOT gate applied on dipole 1 was obtained after 146 iterations with the converged fidelity difference of $1.26\times10^{-6}$; the pulse for dipole 2 took 76 iterations with the converged fidelity of 0.985. In our simulation, the total iteration number is decided by two main effects, the fidelity and the difference between the fidelity and the fidelity of the previous step. If the fidelity is greater than 0.9 and the difference is smaller than $10^{-5}$, the iteration will stop. Both pulses have similar maximum intensity, about 1.2 kV/cm.  In order to verify the performance of this pulse as well as the time evolution of the system within the laser pulse, we chose an initial state $\cos\left(\frac{\pi}{3}\right)\left|01\right\rangle +\sin\left(\frac{\pi}{3}\right)\left|10\right\rangle $ and examined the population evolution.
For an ideal pulse, the final state for the action of a NOT gate on dipole 1 should yield a final state of
$\cos\left(\frac{\pi}{3}\right)\left|11\right\rangle +\sin\left(\frac{\pi}{3}\right)\left|00\right\rangle $.
Correspondingly, the final state for a NOT gate on dipole 2 should be
$\cos\left(\frac{\pi}{3}\right)\left|00\right\rangle +\sin\left(\frac{\pi}{3}\right)\left|11\right\rangle $.
The population evolution due to this pulse assistance is shown in the lower panel of Fig. \ref{NOT}.
Before achieving the final state, the population of each state oscillates a number of cycles. The NOT pulse for dipole 1 produced the converged population for state
$\left|00\right\rangle$ to be 0.756 and that of $\left|11\right\rangle$ to be 0.240, while the pulse for dipole 2 yielded populations of 0.263 and 0.723 for states
$\left|00\right\rangle$ and $\left|11\right\rangle$, respectively. As this final yield result approaches the ideal case $\left(\cos^{2}{60^{o}}=0.75; \; \sin^{2}{60^{o}}=0.25\right)$, we conclude that the optimal laser pulse drives the system from an initial state to a target state according to Table \ref{table_not} for NOT gates.

\begin{figure}
\begin{centering}
\includegraphics[width=0.9\textwidth]{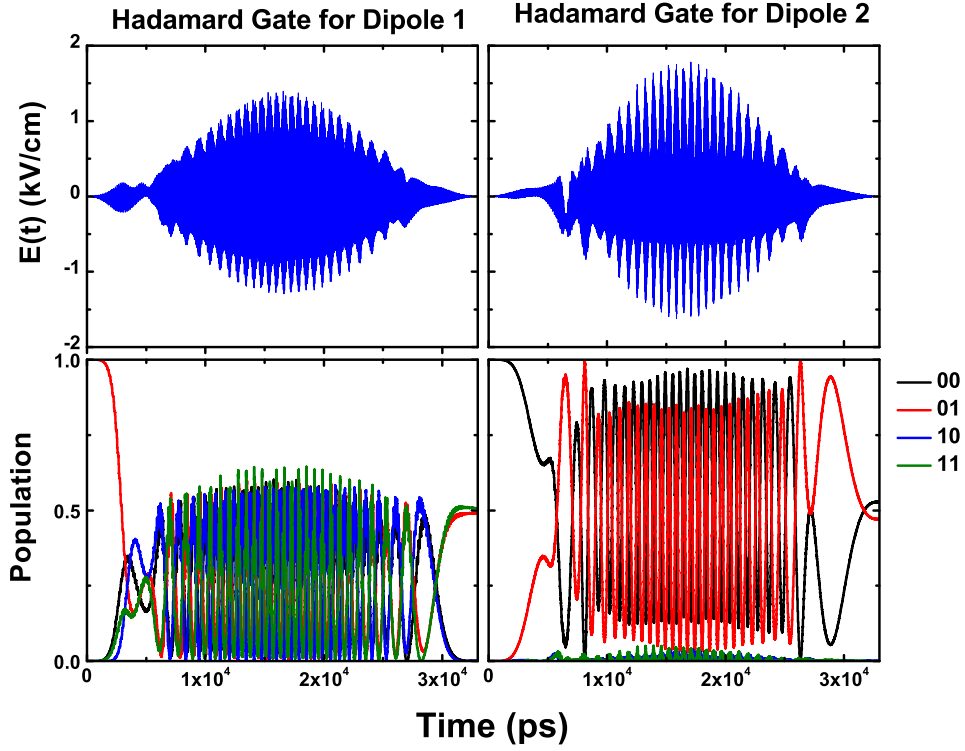}
\par\end{centering}

\caption{Optimized laser pulses for Hadamard gates. The left panel pertains to performing the
Hadamard gate on dipole $1$ and the right one to performing that
on dipole $2$. The lower panels show the population evolution under the Hadamard laser pulse.
In order to exhibit the curves
clearly, we chose $\left|01\right\rangle $ as the initial state;
the final populations of $\left|01\right\rangle $ and $\left|11\right\rangle $
are $0.429$ and $0.566$, respectively. For the case of the Hadamard gate
on dipole $2$, the $\left|00\right\rangle$ qubit was chosen as the initial
state. The populations at the end of the evolution are $0.528$ for
$\left|00\right\rangle $ and $0.470$ for $\left|01\right\rangle $.
The functions of these gates are realized very well by these two pulses.
\label{Hadamard}}
\end{figure}

The converged laser pulses by which to realize the Hadamard gate for both dipoles are shown in Fig. \ref{Hadamard} and the initial and target states are given in Table \ref{table_Hadamard}. The fidelity for each pulse on each site is 0.944 and 0.902, respectively. The maximum intensity for the pulse is around 1.8 kV/cm, slightly larger than that for NOT gates. We selected $\left|00\right\rangle$ as our initial state and plotted the time evolution in Fig. \ref{Hadamard}. The population evolution of the Hadamard gate for dipole 1 is similar to the situation of the NOT gate (Fig. \ref{NOT}) . The population of different states oscillates and finally yields 0.491 for $\left|01\right\rangle$ and 0.508 for $\left|11\right\rangle$. For the Hadamard gate on dipole 2, the population for $\left|10\right\rangle$ and $\left|11\right\rangle$ is almost zero during the entire evolution. But the population switches between $\left|00\right\rangle$ and $\left|01\right\rangle$ . This population oscillation commences at 6 ns and continues until the end of the pulse with a mean value of 0.5 and amplitude around 0.5. The final converged population is 0.529 for $\left|00\right\rangle$ and 0.470 for $\left|01\right\rangle$. The population transfer, shown in Table \ref{table_Hadamard}, going from an initial state to a target state, indicates that the design pulse successfully implemented the Hadamard gate. There is a small variation among all fidelity values we obtained in the simulation, which range from 0.902 to 0.985. This is due to the short duration of the laser pulse we are using in the simulation. The fidelity could be improved further by extending the duration of the laser pulse.

\begin{figure}
\begin{centering}
\includegraphics[width=0.9\textwidth]{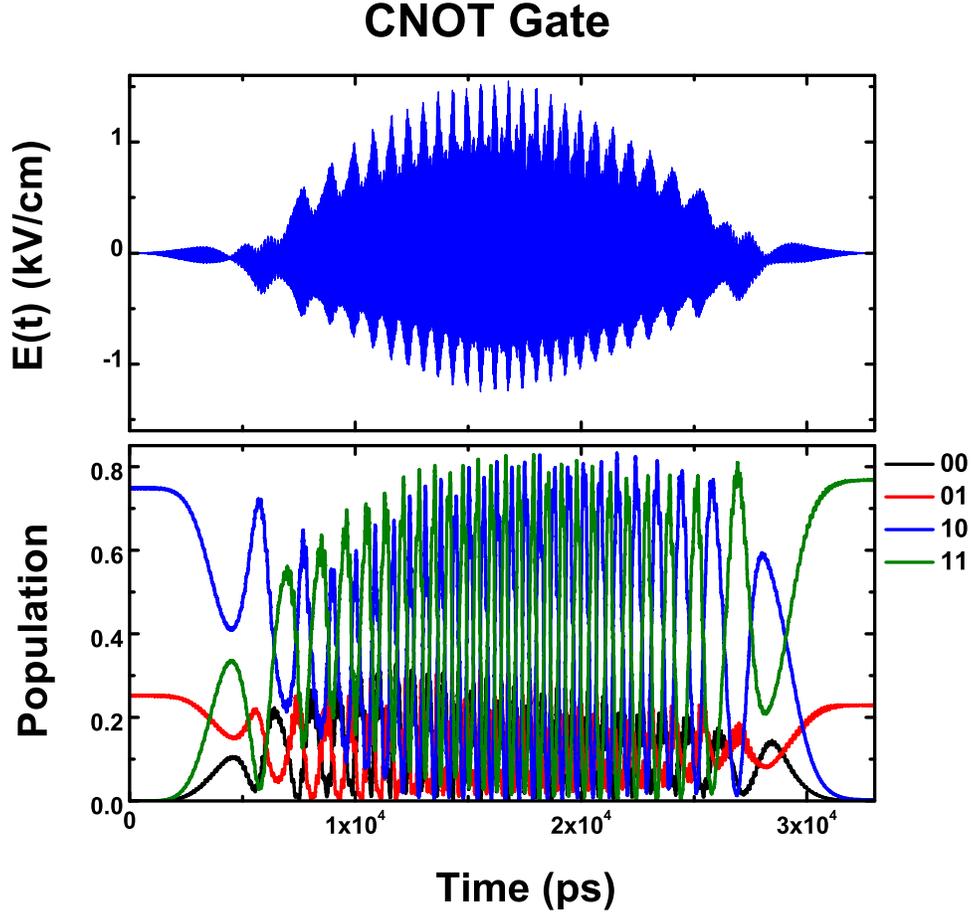}
\par\end{centering}

\caption{Optimized laser pulse for realizing the CNOT gate. The initial and
target states are listed in Table \ref{table_cnot}. In the lower panel,
the population evolution is driven by the CNOT pulse. The initial state
is $\cos\left(\frac{\pi}{3}\right)\left|01\right\rangle +\sin\left(\frac{\pi}{3}\right)\left|10\right\rangle $.
After population oscillations due to the effect of the laser
pulse, the final qubit populations are $0.00363\,\left(\left|00\right\rangle \right)$,
$0.25807\,\left(\left|01\right\rangle \right)$, $0.00273\,\left(\left|10\right\rangle \right)$
and $0.73546\,\left(\left|11\right\rangle \right)$. The populations
of state $\left|10\right\rangle $ and $\left|11\right\rangle$ are
switched, which confirms the correctness of the converged laser pulse.
\label{CNOT}}
\end{figure}

Fig.\ref{CNOT} shows the converged laser pulse which performs the CNOT gate; the pulse is highly oscillatory with total duration time of 33 ns. The rapid oscillation is due to the short system evolution period. The maximum amplitude of the pulse is around 1.5 kV/cm, in the range easily achieved experimentally. The laser pulse was obtained after 441 iterations and yields fidelity of 0.975. If we use the same initial condition as that for NOT gates, the ideal final state should be
$\cos\left(\frac{\pi}{3}\right)\left|01\right\rangle +\sin\left(\frac{\pi}{3}\right)\left|11\right\rangle $.
The simulated population evolution due to the laser pulse evolved is plotted in the lower panel of Fig.\ref{CNOT}.
The final population is 0.0005, 0.2290, 0.0029 and 0.7675 for $\left|00\right\rangle ,\;\left|01\right\rangle ,\;\left|10\right\rangle $ and $\left|11\right\rangle $, respectively.
The population evolution in Fig. \ref{CNOT} follows the initial and target results of Table \ref{table_cnot}, thus confirming the correct operation of the CNOT gate.

\begin{figure}
\begin{centering}
\includegraphics[width=0.9\textwidth]{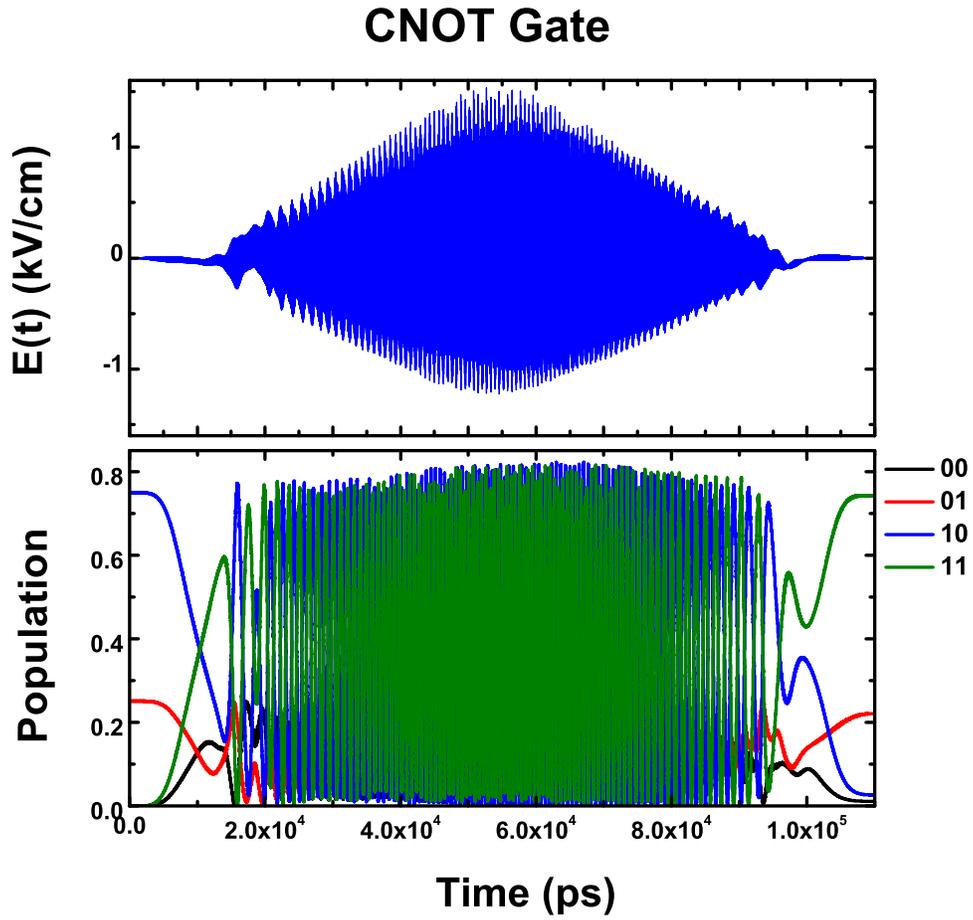}
\par\end{centering}

\caption{Optimized laser pulse for realizing the CNOT gate when the distance between
two dipoles is $75\;\unit{nm}$. The initial and
target states are listed in Table \ref{table_cnot}.
The optimized laser pulse, which is shown in the upper panel, lasts $110\;\unit{ns}$.
The lower panel shows the population evolution driven by the pulse.
The initial state is the same as Fig. \ref{CNOT}.
The final converged populations
for state $\left|01\right\rangle$ and $\left|10\right\rangle$ are $0.22$ and $0.73$, respectively.
\label{CNOT_15P}}
\end{figure}

In order to test, at least modestly, how the MTOCT approach responds to a change in pulse duration for the dipole-dipole system, we increased the spacing between the two dipoles to 75 nm.   The frequency shift then becomes  $\Delta \omega = 14.6 \, \unit{MHz}$  and the duration of the laser pulse is 110 ns.  We carried out the simulations just for the CNOT gate. The optimized laser pulse is shown in Fig. \ref{CNOT_15P}.
After 500 iterations, the converged fidelity is 0.90, which is slightly smaller than for the 50 nm spacing.   Again, we tested one sample initial state. The population evolution is plotted in Fig. \ref{CNOT_15P}.  As before, the population oscillation continued during the whole process and the final population obtained confirms the CNOT operation.  This serves to indicate that the MTOCT approach is stable and provides a useful general means to implement logic gates for dipole-dipole systems.

\section{Discussion}

We have applied the MTOCT methodology to pendular states for a pair of polar molecules (SrO) to determine the optimum laser pulse for implementing the NOT, CNOT and Hadamard quantum logic gates. Our results confirm that, for the conditions adopted ($r_{12}=50\,\unit{nm}$, resulting in $\Delta \omega\,\sim\,50\,\unit{MHz}$), a single laser pulse (with minimum duration $\sim\,33$ ns and amplitude $<2$ kV/cm) suffices to operate these gates with high fidelity. However, computational limitations (storage and time) did not permit us to treat conditions ($r_{12}\,=\,500$ nm, with $\sim50$ kHz) that are considered congenial for experimental implementation.

This shortcoming is also manifest, in different ways, in two previous applications of MTOCT to assess laser-operated logic gates for polar molecules.   Table \ref{table_compare} compares our conditions with those studies. Both nominally emulated the design by DeMille \cite{DeMille2002}, with two ultracold trapped diatomic molecules entangled via dipole-dipole interaction.   The version most akin to ours, presented by Bomble, et al \cite{Bomble2010}, considered a pair of NaCs molecules, with the external static field aligned along the intramolecular axis, $r_{12}$.  Then $\alpha=0^{o}$ (rather than $90^{o}$ as in our case) and hence $\Omega_{\alpha}\,=\,-2\omega$. The qubits were taken as rotational states mixed by the Stark effect to second order (thus a fairly good approximation to the pendular eigenstates we used). The conditions adopted ($r_{12}\,=\,300$ nm, resulting in $\Delta\omega\sim120\,\unit{kHz}$ ) are considered suitable for experimental implementation.
However, in carrying out the MTOCT computations for our system, a very large step size of 10 ps was used.  That avoided entirely the storage and time limitations we encountered (with step size 0.25 ps).   However, we found that replicate calculations using a 10 ps step size gave markedly irreproducible results for the optimal laser properties and population evolution of the logic gates under our simulation system.

The other previous version, presented by Mishima and Yamashita \cite{Yamashita2009b, Yamashita2011} omits altogether an external static electric field, and takes as qubits the lowest "pure" rotational states ($J \, =\,0$, $M\,=\,0$ and $J\,=\,1$, $M\,=\,0$) and lowest vibrational states of each molecule. Specific alignments of the molecular axis with respect to a laboratory fixed z-axis are considered, but that is unrealistic because without an external electric field the molecular axis distribution is isotropic and the laboratory projections of the dipole moments vanish, as emphasized elsewhere \cite{DeMille2002,QiPendular}. Also unrealistic from an experimental perspective is the choice of an extremely small distance between the molecules $\left(r_{12} \,=\, 5\,\unit{nm}\right)$.   That produces large entanglement but would induce severe decoherence \cite{DeMille2002,lattice}. The MTOCT treatment of logic gates is nonetheless of interest, since the choice of "pure" rotational states as qubits results in $\Delta \omega \,=\,0$.  Then, if the molecules are identical, transitions involving qubits on different sites cannot be resolved.  In order to target individually the two sites, two different laser fields were used.  In the MTOCT analysis, the separate laser pulses for gate operations then were much shorter and simpler than in our application using the same laser for both sites. For instance, to perform a CNOT gate using "pure" rotational states with two lasers took only $10^{3}$ ps \cite{Yamashita2009b, Yamashita2011}, whereas our use of pendular qubits with one laser required up to $10^{5}$ ps (see Figs. \ref{CNOT} and \ref{CNOT_15P}).   Although the two-laser mode is theoretically inviting, it requires spatial resolution adequate for each laser to drive only one of the two molecules.  That is not feasible unless the distance between the molecules is much larger than 5 nm.  Another means to contend with $\Delta \omega\,=\,0$, or when it too small to resolve, has been exemplified in designs employing superconducting flux qubits \cite{Groot2010}.  Again, that method requires spatial resolution sufficient to enable qubits on different sites to be driven individually.

Taken together, the three studies of Table \ref{table_compare} illustrate both the utility of MTOCT and the limitations imposed by present-day computational capability.   Aptly, those limitations foster yearning for the arrival of a quantum computer.

\section{ACKNOWLEDGMENTS}
For useful discussions, Jing Zhu thanks Ross Hoehn and Siwei Wei, and appreciates correspondence with Dr Philippe Pellegrini. For support of this work at Purdue, we are grateful to the National Science Foundation CCI center, "Quantum Information for Quantum Chemistry (QIQC)", Award number CHE-1037992, and to the Army Research Office. At Texas A\&M, support was provided by the Institute for Quantum Science and Engineering, as well as the Office of Naval Research and NSF award CHE-0809651.

%\bibliographystyle{aipnum4-1}
%\bibliography{CNOT}

%

\newpage

\begin{longtable}{|c|c|c|c|}
\caption{The initial and target states of the NOT gate. The final fidelities for both NOT gates are 0.967 and 0.985, respectively.}\label{table_not}\endfirsthead
\hline
\multirow{2}{*}{i} & \multirow{2}{*}{Initial State} & \multicolumn{2}{c|}{Target State}\tabularnewline
\cline{3-4}
 &  & NOT gate for Molecule $1$ & NOT gate for Molecule $2$\tabularnewline
\hline
$1$ & $\left|00\right\rangle $ & $\left|10\right\rangle $ & $\left|01\right\rangle $\tabularnewline
\hline
$2$ & $\left|01\right\rangle $ & $\left|11\right\rangle $ & $\left|00\right\rangle $\tabularnewline
\hline
$3$ & $\left|10\right\rangle $ & $\left|00\right\rangle $ & $\left|11\right\rangle $\tabularnewline
\hline
$4$ & $\left|11\right\rangle $ & $\left|01\right\rangle $ & $\left|10\right\rangle $\tabularnewline
\hline
$5$ & $\frac{1}{2}\left(\left|00\right\rangle +\left|01\right\rangle +\left|10\right\rangle +\left|11\right\rangle \right)$ & $\frac{1}{2}\mathbbm{e}^{i\phi}\cdot\left(\left|10\right\rangle +\left|11\right\rangle +\left|00\right\rangle +\left|01\right\rangle \right)$ & $\frac{1}{2}\mathbbm{e}^{i\phi}\cdot\left(\left|01\right\rangle +\left|00\right\rangle +\left|11\right\rangle +\left|10\right\rangle \right)$\tabularnewline
\hline
\end{longtable}

\vspace{24pt}

\begin{longtable}{|c|c|c|c|}
\caption{The initial and target states of the Hadmard gate. The yield fidelities are 0.944 and 0.902}\label{table_Hadamard}\endfirsthead
\hline
\multirow{2}{*}{i} & \multirow{2}{*}{Initial State} & \multicolumn{2}{c|}{Target State}\tabularnewline
\cline{3-4}
 &  & Hadamard gate for Molecule $1$ & Hadamard gate for Molecule $2$\tabularnewline
\hline
$1$ & $\left|00\right\rangle $ & $\frac{1}{\sqrt{2}}\left(\left|00\right\rangle +\left|10\right\rangle \right)$ & $\frac{1}{\sqrt{2}}\left(\left|00\right\rangle +\left|01\right\rangle \right)$\tabularnewline
\hline
$2$ & $\left|01\right\rangle $ & $\frac{1}{\sqrt{2}}\left(\left|01\right\rangle +\left|11\right\rangle \right)$ & $\frac{1}{\sqrt{2}}\left(\left|00\right\rangle -\left|01\right\rangle \right)$\tabularnewline
\hline
$3$ & $\left|10\right\rangle $ & $\frac{1}{\sqrt{2}}\left(\left|00\right\rangle -\left|10\right\rangle \right)$ & $\frac{1}{\sqrt{2}}\left(\left|10\right\rangle +\left|11\right\rangle \right)$\tabularnewline
\hline
$4$ & $\left|11\right\rangle $ & $\frac{1}{\sqrt{2}}\left(\left|01\right\rangle -\left|11\right\rangle \right)$ & $\frac{1}{\sqrt{2}}\left(\left|10\right\rangle -\left|11\right\rangle \right)$\tabularnewline
\hline
$5$ & $\frac{1}{2}\left(\left|00\right\rangle +\left|01\right\rangle +\left|10\right\rangle +\left|11\right\rangle \right)$ & $\frac{1}{\sqrt{2}}\mathbbm{e}^{i\phi}\cdot\left(\left|00\right\rangle +\left|01\right\rangle \right)$ & $\frac{1}{\sqrt{2}}\mathbbm{e}^{i\phi}\cdot\left(\left|00\right\rangle +\left|10\right\rangle \right)$\tabularnewline
\hline
\end{longtable}

\vspace{24pt}

\begin{longtable}{|c|c|c|}
\caption{The initial and target states of the CNOT gate. The converged fidelity is 0.975.}\label{table_cnot}\endfirsthead
\hline
i & Initial State & Target State\tabularnewline
\hline
$1$ & $\left|00\right\rangle $ & $\left|00\right\rangle $\tabularnewline
\hline
$2$ & $\left|01\right\rangle $ & $\left|01\right\rangle $\tabularnewline
\hline
$3$ & $\left|10\right\rangle $ & $\left|11\right\rangle $\tabularnewline
\hline
$4$ & $\left|11\right\rangle $ & $\left|10\right\rangle $\tabularnewline
\hline
$5$ & $\frac{1}{2}\left(\left|00\right\rangle +\left|01\right\rangle +\left|10\right\rangle +\left|11\right\rangle \right)$ & $\frac{1}{2}\mathbbm{e}^{i\phi}\cdot\left(\left|00\right\rangle +\left|01\right\rangle +\left|11\right\rangle +\left|10\right\rangle \right)$\tabularnewline
\hline
\end{longtable}

\newpage

\begin{longtable}{c|ccccccccc}
\caption{Comparison with other polar molecular systems}\label{table_compare}\endfirsthead
\hline
 & Molecule & $\mu\left(D\right)$ & $R\left(\unit{nm}\right)$ & $B\left(\unit{cm^{-1}}\right)$ & $\nicefrac{\Omega}{B}$ & $x$ & $x^{\prime}$ & $\Delta\omega\left(\unit{MHz}\right)$ & $\tau\left(\unit{ns}\right)$\tabularnewline

\hline
Ref. \cite{Bomble2010}  & NaCs & 4.6 & 300 & 0.059 & $6.7\times10^{-5}$ & $\;\;2.0\;$ & 2.6 & 0.12 & $1.2\times10^{4}$\tabularnewline
This paper & SrO & 8.9 & 50 & 0.33 & $9.6\times10^{-3}$ & $\;\;2\;$ & 3 & 51 & 33\tabularnewline
Ref. \cite{Yamashita2009b,Yamashita2011} & NaCl & 8.4 & 5 & 0.22 & 13 & - & - & 0 & 2.6\tabularnewline
Ref. \cite{Yamashita2009b} & NaBr & 8.2 & 5 & 0.15 & 18 & - & - & 0 & -\tabularnewline
\hline
\end{longtable}

\newpage

\end{document}